\documentclass[%
onecolumn,
superscriptaddress,
 amsmath,amssymb,
 aps,prfluids,longbibliography
]{revtex4-2}

\usepackage{graphicx}
\usepackage{amsmath}
\usepackage{bm}
\usepackage{gensymb}
\usepackage{float}

\usepackage[normalem]{ulem}

\usepackage{hyperref}
\usepackage[svgnames]{xcolor}
\hypersetup{colorlinks=true, citecolor=blue, urlcolor=blue, linkcolor=blue}

\begin{document}


\title{Tracking the rotation of light magnetic particles in turbulence}


\author{Chunlai Wu}
\affiliation{Fluids and Flows group and J.M. Burgers Center for Fluid Mechanics,
Department of Applied Physics and Science Education,
Eindhoven University of Technology, 5600 MB Eindhoven, Netherlands}

\author{Rudie P. J. Kunnen}
\affiliation{Fluids and Flows group and J.M. Burgers Center for Fluid Mechanics,
Department of Applied Physics and Science Education,
Eindhoven University of Technology, 5600 MB Eindhoven, Netherlands}

\author{Ziqi Wang}
\affiliation{Fluids and Flows group and J.M. Burgers Center for Fluid Mechanics,
Department of Applied Physics and Science Education,
Eindhoven University of Technology, 5600 MB Eindhoven, Netherlands}

\author{\mbox{Xander M. de Wit}}
\affiliation{Fluids and Flows group and J.M. Burgers Center for Fluid Mechanics,
Department of Applied Physics and Science Education,
Eindhoven University of Technology, 5600 MB Eindhoven, Netherlands}

\author{Federico Toschi}
\affiliation{Fluids and Flows group and J.M. Burgers Center for Fluid Mechanics,
Department of Applied Physics and Science Education,
Eindhoven University of Technology, 5600 MB Eindhoven, Netherlands}
\affiliation{Consiglio Nazionale delle Ricerche - Istituto per le Applicazioni del Calcolo, Rome I-00185, Italy}

\author{Herman J. H. Clercx}
\email{h.j.h.clercx@tue.nl}
\affiliation{Fluids and Flows group and J.M. Burgers Center for Fluid Mechanics,
Department of Applied Physics and Science Education,
Eindhoven University of Technology, 5600 MB Eindhoven, Netherlands}


\begin{abstract}
Particle-laden turbulence involves complex interactions between the dispersed and continuous phases. Given that particles can exhibit a wide range of properties, such as varying density, size, and shape, their interplay with the flow can lead to various modifications of the turbulence. Therefore, understanding the dynamics of particles is a necessary first step toward revealing the behavior of the multiphase system. Within the context of particle dynamics, accurately resolving rotational motion presents a significantly greater challenge compared to translational motion. We propose an experimental method to track the rotational motion of spherical, light, and magnetic particles with sizes significantly smaller than the Taylor microscale, typically an order of magnitude larger than the Kolmogorov scale of the turbulence in which they are suspended. The method fully resolves all three components of the particle angular velocity using only 2D images acquired from a single camera. This technique enables a detailed investigation of the rotational dynamics of magnetic particles subjected simultaneously to small-scale turbulent structures and external magnetic forcing. Beyond advancing the study of particle dynamics in turbulence, this approach opens new possibilities for actively modulating turbulence through externally applied magnetic fields.
\end{abstract}

\begin{abstract}
    Particle-laden turbulence involves highly complex interactions between the dynamics of the dispersed particle phase and that of the continuous turbulent phase. Particles with different physical properties, such as shape, density and size, are particularly interesting as these can be sensitive to or influence different specific aspects of the flow. In this contribution we focus on the coupling between the angular dynamics of particles and turbulent vorticity in turbulence. To this end, we detail the development and demonstrate the capabilities of a special experimental setup where the angular dynamics of particles can be both measured (via optical image processing techniques) and imposed (via the application of external rotating magnetic fields). We further discuss how particles of different densities and magnetic properties can be manufactured, with specific interest towards light particles as these can be transformed into vorticity probes and localized forcing acting on small-scale vortex filaments. The ordinary optical method for particle tracking velocimetry is here extended to accurately measure the angular velocity of particles with sizes significantly smaller than the Taylor microscale, typically an order of magnitude larger than the Kolmogorov scale of the turbulence in which they are suspended. Remarkably, the optical method employed in this study enables accurate measurement of the full three-dimensional angular velocity vector of particles using only two-dimensional image sequences, thereby requiring only one camera. The present contribution discusses in detail how the different design choices are entangled with each other and which considerations are relevant when designing such an experimental setup. Beyond advancing the study of particle dynamics in turbulence, this approach opens up possibilities for actively studying how to modulate turbulence through externally applied magnetic fields.
\end{abstract}

\maketitle

\section{Introduction} \label{sec:intro}
Turbulent flows laden with particles are ubiquitous in both natural and engineering environments. While the dynamics of particles immersed in such flows are governed by a large number of parameters, the complex multi-way coupling between the dispersed and continuous phases also presents opportunities to modify the flow properties themselves \cite{Brandt2022, Mathai2020}. Owing to the effect of local pressure, particles heavier than the fluid tend to accumulate in regions with high strain rates, whereas lighter-than-fluid particles are preferentially drawn into regions where vorticity dominates over strain \cite{Calzavarini2008, Balachandar2010}. Light particles have been the subject of extensive research in recent decades, particularly in studies on the dynamics of light particles and their modulation of turbulence \cite{Volk2008, vanAartrijk2010, Mercado2012, MathaiCalzavarini2016, Oka2021, Wang2002, Motoori2023, Wang2024, Yeo2010, Aliseda2011}. With the rapid advancement of Lagrangian Particle Tracking (LPT), the translational motion of particles can now be experimentally resolved at sub-pixel resolution ($\sim 0.1\, \text{pixel}$) \cite{Schroeder2023}. However, accurately tracking the rotational motion of particles remains a significant challenge. One effective approach involves applying black-and-white surface patterns to particles, enabling reconstruction of their orientations \cite{Mathai2016, Zimmermann2011}. Grafting the surface of a large polymer particle (with a diameter of approximately 100$\eta$) with six to eight fluorescent tracer particles enables the recovery of the particle's center and radius, thereby allowing for the determination of both its position and rotational motion \cite{Klein2013}. These techniques, however, typically require predefined surface patterns or marker distributions and are applicable only to spherical particles that are substantially larger than the Kolmogorov scale of turbulence, often limiting the method to a small number of particles. Various rotational tracking methods have also been developed for particles with spheroidal and anisotropic shapes, such as fibers and crosses \cite{Cole2016, Marcus2014, Ibarra2023}. From the reverse perspective, considerable numerical and experimental work has focused on how particles modulate turbulence. For example, the injection of microbubbles has shown significant potential in altering turbulence structures and reducing drag \cite{GutierrezTorres2008, Rawat2019}. Recent studies on particle-laden Taylor–Couette flows have revealed how particle-induced rotational dynamics, including inertia and hydrodynamic interactions, can lead to complex phenomena such as hysteresis, ribbon instabilities, and flow transitions \cite{Kang2021, Kang2023, Kang2024}. In homogeneous shear turbulence, the turbulent kinetic energy exhibits a non-monotonic dependence on the concentration of neutrally buoyant particles: initially, turbulence is increasingly attenuated as particle concentration rises, but beyond a certain threshold, further increases in concentration lead to an enhancement of turbulent kinetic energy \cite{Yousefi2020}. 

In this study, we present an experimental method that fully resolves the rotational motion of light spherical particles with diameters $D_{\text{p}}$ within the range $\eta < D_{\text{p}} < \lambda$, where $\eta$ denotes the Kolmogorov length scale and $\lambda$ the Taylor microscale. The tracking method is developed based on an existing three-dimensional (3D) rotation tracking algorithm that requires only two-dimensional (2D) images acquired from a single camera \cite{Niggel2023}. A key advantage of these particles is their buoyancy: being lighter than water, they naturally serve as an intermediate medium for probing and modulating turbulence from within vortex-dominated regions. To enable external control, a thin magnetic coating is applied to the particles, granting them magnetic properties. This allows them to be actuated by an externally applied rotating magnetic field, even while immersed in a turbulent flow. This experimental methodology has been implemented in a newly designed setup that simultaneously applies turbulent and magnetic forcing to the particles. The system opens a new avenue for manipulating the dynamics of particles suspended in turbulence and for modulating turbulence itself at the smallest flow scales through external forcing.

This paper is structured as follows: the design and validation of the experimental setup are discussed in Sec. \ref{sec:setup}. The fabrication process of the light magnetic particles is described in Sec. \ref{sec:magnetic particles}. Sec. \ref{sec:Code} details the image processing and rotation tracking algorithm, along with its validation in terms of tracking accuracy under varying rotation frequencies, image resolutions, and focus levels. Sec. \ref{sec:experiments} illustrates our method with experimental results of particle rotation tracking in turbulent flow under the influence of rotating magnetic fields at several different frequencies. Finally, Sec. \ref{sec: Conclusion} summarizes our main findings and provides a brief discussion of our method.

\section{Experimental Setup} \label{sec:setup}
The experimental setup is designed to generate a turbulent flow and a rotating magnetic field within the measurement volume. In this system, a Von Kármán flow is produced inside an octagonal prism water tank by two coaxially counter-rotating disks mounted at the top and bottom. The two disks are identical in geometry and consistently operate at the same rotation speed. Two pairs of Helmholtz coils, positioned horizontally and vertically around the water tank, function as the planar rotating magnetic field generator.

One of the primary objectives of this setup is to manipulate the rotational motion of magnetic particles by adjusting both the turbulence intensity and the magnetic interaction between the particles and the rotating magnetic field. Therefore, obtaining a preliminary estimate of the turbulent flow characteristics, the magnetic field properties, and the attributes of the magnetic particles is essential to ensure that the experimental parameter space covers a sufficiently broad regime where the rotational dynamics of the magnetic particles is not dominated solely by either field. Subsequently, a carefully integrated design process was carried out to balance multiple trade-offs in parameters, ultimately resulting in the final configuration of the current facility. For example, reducing the radius of the coils enhances the magnetic field strength, which benefits magnetic control of the particles. However, this comes at the cost of reducing the available space for turbulence generation, which leads to a smaller Kolmogorov scale, as the flow must remain fully turbulent with the integral scale Reynolds number $\text{Re} = 2\pi R_{\text{d}}^2\Omega/\nu \geq 3300$ \cite{Ravelet2008}, where $R_{\text{d}}$ and $\Omega$ are the radius and the rotation frequency of the two rotating disks, respectively, and $\nu = 10^{-6}$ m\textsuperscript{2}/s is the kinematic viscosity of water. As a result, smaller magnetic particles are required to effectively interact with the smallest turbulent flow structures, which in turn poses challenges for particle fabrication, imaging resolution, and magnetic responsiveness. Therefore, the final design represents a compromise that sufficiently satisfies the requirements of both turbulence generation and magnetic field control to ensure the desired experimental conditions. In the following sections, we introduce and discuss each subsystem along with its validation, gradually building up to the full experimental setup.

\subsection{ Turbulence generator: the French washing machine}
\begin{figure}[htbp]
    \centering
    \includegraphics[width=0.75\textwidth, angle=0]{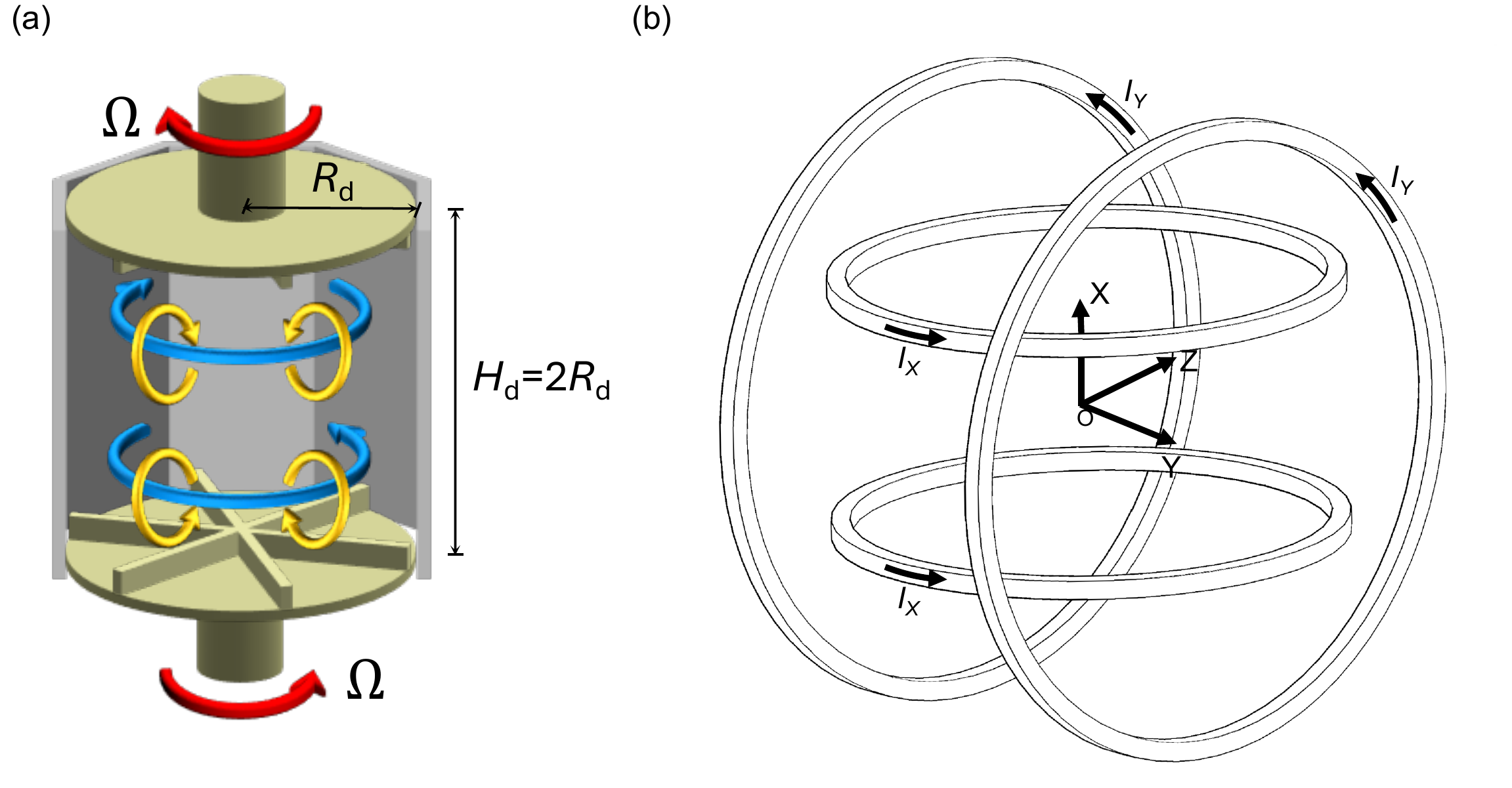} 
    \caption{(a) The turbulence generator, commonly referred to as the French washing machine, is shown with the primary and secondary flow directions indicated by blue and yellow arrows, respectively. The radius of each of the two identical impellers is represented by $R_\text{d}$, and they are separated by a distance of $H_\text{d} = 2R_\text{d}$. The impellers rotate in opposite directions at a frequency $\Omega$. (b) Two pairs of Helmholtz coils are mounted perpendicularly, with their axes aligned along the \textit{x} and \textit{y} axes, respectively. The amplitudes of the alternating currents in the two coil pairs, $I_\text{X}$ and $I_\text{Y}$, are slightly adjusted to ensure that the magnetic flux densities $\bm{B_x}$ and $\bm{B_y}$ have equal magnitudes at the origin $\bm{O}$, compensating for the difference in coil radii. A 90° phase difference between $I_\text{X}$ and $I_\text{Y}$ results in a rotating planar magnetic field in the \textit{xy} plane. }
    \label{fig:FWM_coils}
\end{figure}

The turbulent flow generated by two counter-rotating disks equipped with either straight or curved blades and enclosed within a cylindrical or polygonal prism is known as Von Kármán flow \cite{Zandbergen1987, Poncet2008}. This type of experimental setup is conventionally referred to as the ``French washing machine'' and is capable of producing a highly intense turbulent flow  at its center. The schematic representation of the mean large-scale flows is shown in Fig. \ref{fig:FWM_coils}a. In this setup, the two identical impellers have a diameter of $R_{\text{d}} = 75$ mm, and are separated by a distance $H_{\text{d}} = 2R_{\text{d}} = 150$ mm. Each impeller is driven by a brushless DC motor (Maxon EC-i 30), operating at a constant angular speed $\Omega$ throughout the experiments.   
To estimate beforehand the characteristic properties of the Von Kármán flow using only the global parameters mentioned above, obtaining the dissipation rate $\varepsilon$ is of primary importance. We introduce an empirical constant $C_{\varepsilon}$ for the estimation of $\varepsilon$, which subsequently enables the calculation of Kolmogorov scales.

From dimensionless analysis it is shown that the required torque $\Gamma$ for driving the two impellers can be written as 
\begin{equation}
    \Gamma = \rho R_{\text{d}}^5 \Omega^2 f(\text{Re}),
\end{equation}
where $\rho = 1000$ kg/m\textsuperscript{3} is the density of water, $f(\text{Re})$ is an unknown function of the integral scale Reynolds number $\text{Re}$. In the simplest case of rotating disks fitted with blades, the function $f(\text{Re})$ is approximately constant, meaning it is independent of $\text{Re}$, and depends only on the geometric parameters of the setup, such as the shape and number of the blades \cite{Mordant1997, Labbe1996}. This torque $\Gamma$ can be obtained from the total input power $P = 2\pi\Omega\Gamma$ that is consumed by the disks and transferred to the turbulence.

The turbulence dissipation rate $\varepsilon$ is given by

\begin{equation}
\varepsilon = \frac{P}{\rho V_\text{w}} = C_\varepsilon \left( \frac{R_{\text{d}}^3 \Omega^3}{H_{\text{d}}} \right),
\label{eq:epsilon}
\end{equation}
where $C_\varepsilon$ is a constant that has taken $f(\text{Re})$ into account, and $V_\text{w} \approx \pi R_{\text{d}}^2H_{\text{d}}$ is the volume of the water tank. By collecting experimental data from the literature ~\citep{Volk2008, Volk2011, Mordant2004, Voth1998, Voth2002, Huck}, the average value of $C_\varepsilon$ can be obtained using

\begin{equation}
C_\varepsilon = \frac{\nu^3}{\eta^4} \left( \frac{H_{\text{d}}}{R_{\text{d}}^3 \Omega^3} \right),
\label{eq:C_epsilon}
\end{equation}
where $\eta$ is the Kolmogorov length scale measured or calculated in the experiments in the above literature. In those experiments, the working fluid is water. Finally, we have an average value of $C_\varepsilon \approx 11$. 

Given the constant $C_\varepsilon$ and other global physical parameters of the French washing machine, the Kolmogorov length and time scales follow as $\eta = \left(\nu^3/\varepsilon\right)^\frac{1}{4}$ and $\tau_\eta = \left(\nu/\varepsilon\right)^\frac{1}{2}$, respectively. The Taylor microscale $\lambda $ can be calculated by $\lambda = \sqrt{15} \tau_{\eta} u'$, where $u' = (\varepsilon L)^{1/3}$ is the root-mean-square velocity and $L = R_{\text{d}}$ represents the integral length scale. In our experiments, we aim to generate turbulence such that the particle size satisfies $\eta < D_{\text{p}} < \lambda$, ensuring sensitivity to the small-scale structures of the flow. To achieve this, we target a Taylor Reynolds number of $\text{Re}_{\lambda} = \sqrt{15u'L/\nu} \approx 350$,  which is a typical value reported for similar devices \cite{Andreas2023}. Key parameters and features of the setup are discussed in Sec. \ref{sec:Integrated design}.

\subsection{Magnetic field generator: Helmholtz coils}
A pair of Helmholtz coils, shown in Fig. \ref{fig:FWM_coils}b, is a conventional device that is widely used to generate a homogeneous one-dimensional magnetic field in a restricted volume between both coils. Two identical circular coils are separated along their axis by a distance equal to the coil radius $R_{\text{c}}$. When identical currents $I$ flow in the same direction in the two coils, such a homogeneous magnetic field is produced. Similarly to turbulence generated in the French washing machine, Helmholtz coils cannot provide a homogeneous field in the whole region surrounded by them but only in a delimited space close to the centroid (the origin O in Fig. \ref{fig:FWM_coils}b). The magnetic field homogeneity that describes the variability of the magnetic field flux density $\bm{B}$ is usually expressed in a percentage and defined as follows \cite{Wang2002}:  
\begin{equation}
     H[\%] = \frac{\left| \bm{B} - \bm{B_O} \right|}{\left| \bm{B_O} \right|} \times 100\% ,
\end{equation}
while the magnetic field flux density at the center point can be calculated by
\begin{equation}
   |\bm{B_O}| = \left(\frac{4}{5}\right)^{\frac{3}{2}}\frac{\mu_0 nI}{R_{\text{c}}},
   \label{eq:Magnetic flux density}
\end{equation}
where $\mu_0 = 4\pi \times10^{-7}$ N/A\textsuperscript{2} is the magnetic permeability of vacuum, $n$ is the number of windings in each coil, and $I$ is the current magnitude. When the distance to the center is less than $0.1R_{\text{c}}$, in any direction, we have $H<1\%$ \cite{Restrepo2017}.  

In this setup, two pairs of Helmholtz coils are mounted perpendicularly along the \textit{x} and \textit{y} axes, carrying alternating currents with a phase difference of $\pi/2$ to generate a rotating magnetic field in the \textit{xy} plane. Since the radii of the two coil pairs differ to facilitate assembly, the number of windings $n$ and the magnitude of the current are adjusted accordingly (while remaining identical within each pair) to ensure that both pairs generate magnetic flux densities of equal magnitude.

\subsection{Integrated design of the setup} \label{sec:Integrated design}

To generate a homogeneous turbulent flow and a rotating planar magnetic field within the measurement volume, two pairs of Helmholtz coils are mounted closely around the octagonal prism water tank in horizontal and vertical orientations. Consequently, the dimensions of the water tank and coils are strongly interdependent. Given that $\varepsilon = C_\varepsilon \left( R_{\text{d}} \Omega\right)^3/H_{\text{d}}$ and $H_{\text{d}} = 2R_{\text{d}}$, the Kolmogorov length scale becomes $\eta = \left(\nu^3H_{\text{d}}/\left(C_{\varepsilon}R_{\text{d}}^3\Omega^3\right)\right)^{1/4} $. Keeping the integral scale Reynolds number $\text{Re} = 2\pi R_{\text{d}}^2\Omega/\nu$ constant implies $\Omega \propto 1/R_{\text{d}}^2$, and therefore, $\eta \propto  R_{\text{d}}$. This shows that $\eta$ increases as the size of the water tank, represented by the impeller radius $ R_{\text{d}} $, increases. On the other hand, $|\bm{B_O}| = \left(4/5\right)^{3/2}\mu_0 nI/R_{\text{c}}$ indicates that the magnetic flux density of the field is inversely proportional to the coil radius $ R_{\text{c}} $. The magnetic particles in our experiments should have a size compatible with the dissipation range of the turbulent flow. The magnetic flux density of the external magnetic field should be large enough to enable actuation of the magnetic particles. As the Kolmogorov length scale $\eta$ increases with $R_\text{d}$ and the magnetic flux density $|\bm{B}|$ in the measurement volume increases with decreasing $R_\text{c}$, which implies decreasing $R_\text{d}$, an optimal set of ($R_\text{d}$, $R_\text{c}$) needs to be determined depending on the size of the magnetic particles. 

Choosing an appropriate entry point for parameter design is essential. Here we set the target to be the Taylor Reynolds number of the turbulence, with $\text{Re}_{\lambda} \approx 350$. By combining this with Eq. (\ref{eq:epsilon}), we determine that the impellers must rotate at a frequency $\Omega \geq 0.83$ Hz to achieve the desired turbulence intensity, given an impeller radius of $R_{\text{d}} = 75$ mm. Under these conditions, the estimated dissipation rate is $\varepsilon = 0.018$ m\textsuperscript{2}/s\textsuperscript{3}. The estimated Kolmogorov length and time scales are $\eta = 0.087$ mm and $\tau_{\eta} = 7.56$ ms, respectively. The corresponding integral scale Reynolds number is $\text{Re} \approx 29000 \gg 3300$, indicating fully developed turbulence.

The primary torques governing the rotational motion of the magnetic particles are the viscous torque, here approximated by the Stokes torque, and the magnetic torque. For effective magnetic control, the magnetic torque must exceed the typical Stokes torque. These torques can be quantified as follows \cite{Brenner1963, Abbott2007}:

\begin{subequations}
\begin{align}
    |\bm{T_\text{St}}| &= \pi\rho\nu D_{\text{p}}^3 \delta\omega, 
    \label{eq:Stokes torque}\\
    |\bm{T_\text{m}}|_{max} &= \frac{1}{\mu_0} \chi_V V_{\text{p}} {|\bm{B}|}^2, 
    \label{eq:Magnetic torque}
\end{align}
\end{subequations}
where $\delta\omega$ denotes the angular velocity of the surrounding flow on the magnetic particles, $ \chi_V $ is the average volume magnetic susceptibility of a magnetic particle, and $ V_{\text{p}} = \pi D_{\text{p}}^3/6 $ represents the volume of a magnetic particle. Assuming the magnetic particle diameter $ D_{\text{p}} $ is on the order of the dissipative scale, $ \eta < D_{\text{p}} < \lambda$, with the Taylor microscale $\lambda = \sqrt{15} \tau_{\eta}u'  = \sqrt{15} \tau_{\eta} \left(\varepsilon R_\text{d} \right)^{1/3} \approx 3.2$ mm, we take $D_{\text{p}} \approx 10\eta = 0.87$ mm. To estimate $\delta\omega$ in the range of $r \approx 10 \eta$, we define $\delta\omega \equiv d(\sqrt{D_{NN}}) / dr$, where $D_{NN}$ is the second-order transverse Eulerian structure function, representing the transverse velocity difference between two points in the flow field separated by a distance $r$ \cite{Pope}. In the inertial range ($r \gg \eta$) it becomes $D_{NN} = 4C_2(\varepsilon r)^{2/3}/3$, where $C_2 = 2.1$ \cite{Gibert2010}. Using these expressions, we have $\delta\omega \geq 16$ s\textsuperscript{-1} for $r = D_{\text{p}} \approx 10\eta$. Assuming a typical value for the average magnetic susceptibility of sub-millimeter particles exhibiting super-paramagnetism \cite{Shanko2021}, we take $\chi_V = 0.05$. Then the minimum magnetic field flux density required to balance the Stokes torque is calculated as $|\bm{B}|_{min} = \sqrt{6\rho\nu\mu_0\delta\omega/\chi_V} = 1.5$ mT for $\delta\omega = 16$ s\textsuperscript{-1}. For the convenience of estimation, we assume the magnetism of the particles to be isotropic. However, in reality, it is precisely the presence of magnetic anisotropy that enables the particles to continuously follow the rotation of the external magnetic field. A detailed explanation of this mechanism is provided in Sec. \ref{sec:magnetic particles}. Appendix \ref{Appendix1} provides a comprehensive summary of all the equations used in the design of the setup. 

Considering the necessary support structure for the coils and the required clearance between the two pairs, a reasonable estimate for $R_{\text{c}}$ is 150 mm. To meet the lower bound of $|\bm{B}| = 1.5$ mT, as given by Eq. (\ref{eq:Magnetic flux density}), the required value of $nI$ is 250 A, which corresponds, for example, to 25 turns of wire carrying a 10 A current. 

\begin{figure}[htbp]
    \centering
    \includegraphics[width=1\textwidth, angle=0]{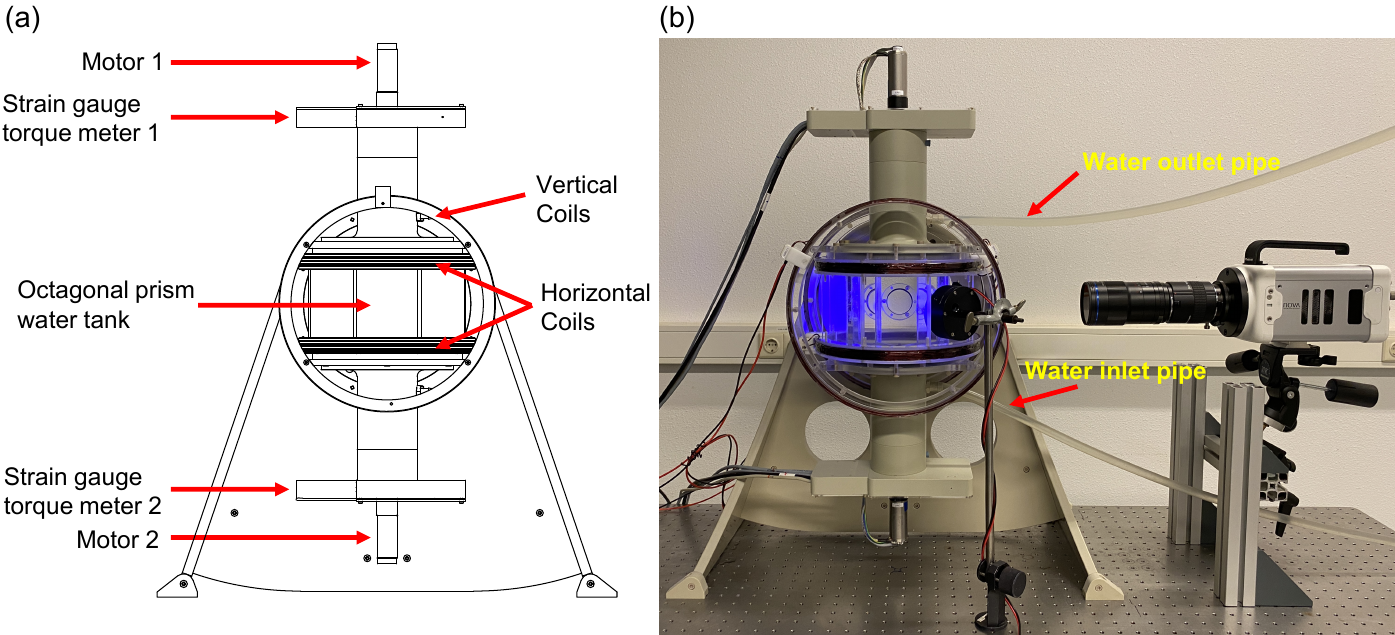} 
    \caption{(a) Schematic drawing of the experimental setup. The radius of each impeller is $R_\text{d} = 75$ mm, and the axial separation between the two impellers is $H_\text{d} = 150$ mm. Two torque meters are used to measure the energy input from the motors to the water to calculate the turbulence dissipation rate. The vertical coils, aligned along the \textit{x} axis, have a radius of 150 mm and consist of 34 wire turns. The horizontal coils, aligned along the \textit{y} axis, have a radius of 125 mm with 28 turns. Together, the two coil pairs generate a planar rotating magnetic field with a magnitude of 1.6 mT. (b) Photograph of the experimental setup. The water tank is filled from the bottom through an inlet pipe and drained from the top through an outlet pipe.}
    \label{fig:Setup}
\end{figure}

In the end, the general configuration of the setup, as shown in Fig. \ref{fig:Setup}, consists of two counter-rotating impellers with a radius of $ R_{\text{d}} = 75 $ mm, each equipped with eight straight blades of 10 mm in height. The separation distance between the flat surfaces of the two impellers is $H_{\text{d}} = 150$ mm. The apothem of the octagonal water tank's cross-section is 75.1 mm.  

The measured dissipation rate $ \varepsilon $ is determined from the driving torque on the two impellers as $\varepsilon = T \Omega/(\rho V_\text{w})$, where $ T $ is the sum of the driving torques from the impellers at the top and bottom of the water tank and $ V_\text{w} = 2.8 \times 10^{-3} $ m\textsuperscript{3} is the volume of the water tank. The driving torques are measured using two strain gauge torque meters mounted on the motors. When the impellers rotate at 0.83 Hz, the dissipation rate is $ \varepsilon = 0.037 \pm 0.002 $ m\textsuperscript{2}/s\textsuperscript{3}, corresponding to a Taylor Reynolds number of $\text{Re}_\lambda = 398$, a Kolmogorov length scale of $ \eta = 0.072 $ mm, and a Kolmogorov time scale of $ \tau_{\eta} = 5.20 $ ms. Compared to the estimated values of $\text{Re}_\lambda = 350$, $ \eta = 0.087 $ mm, and $ \tau_{\eta} = 7.56 $ ms, the empirical estimate demonstrates acceptable accuracy for the preliminary design of the setup.  

For ease of assembly, the two pairs of Helmholtz coils have different radii, measuring 150 mm and 125 mm, respectively. Each pair of coils consists of 34 and 28 turns of wire carrying an alternating current of up to 8 A. Due to differences in the cross-sectional size of the two coil pairs, electrical resistance, and manufacturing tolerances, the actual operating current in each pair of coils may differ slightly to ensure that the value of $nI/R_c$ remains equal for both pairs. The rotating planar magnetic field is calibrated using a SENIS F3A magnetic-field-to-voltage transducer equipped with a fully integrated 3D Hall probe. Within the frequency range of 1 to 50 Hz, the maximum magnetic flux density remains stable at $ 1.60 \pm 0.02 $ mT, which meets the 1.5 mT requirement necessary to balance the hydrodynamic drag torque exerted on the magnetic particles.

\section{Light magnetic particles} \label{sec:magnetic particles}
The particles used in our experiments possess three key characteristics that enable the study of their rotational motion under the combined influence of turbulence and magnetic interactions with the rotating magnetic field. The particle diameter is in the range between the Kolmogorov scale and the Taylor microscale, with $D_\text{p} \approx 10\eta$, which ensures interaction with the smallest turbulent flow structures. The particles have sufficiently anisotropic magnetic susceptibility, providing a strong magnetic torque capable of overcoming the turbulent torque. Finally, the density is much lower than that of water, allowing them to accumulate in high-vorticity regions.

Magnetic materials are generally much denser than water. Therefore, to achieve an overall particle density lower than that of water, a significantly lighter material must serve as the primary component of the magnetic particles to compensate for this density disparity. 

\begin{figure}[H]
    \centering
    \includegraphics[width=0.85\textwidth, angle=0]{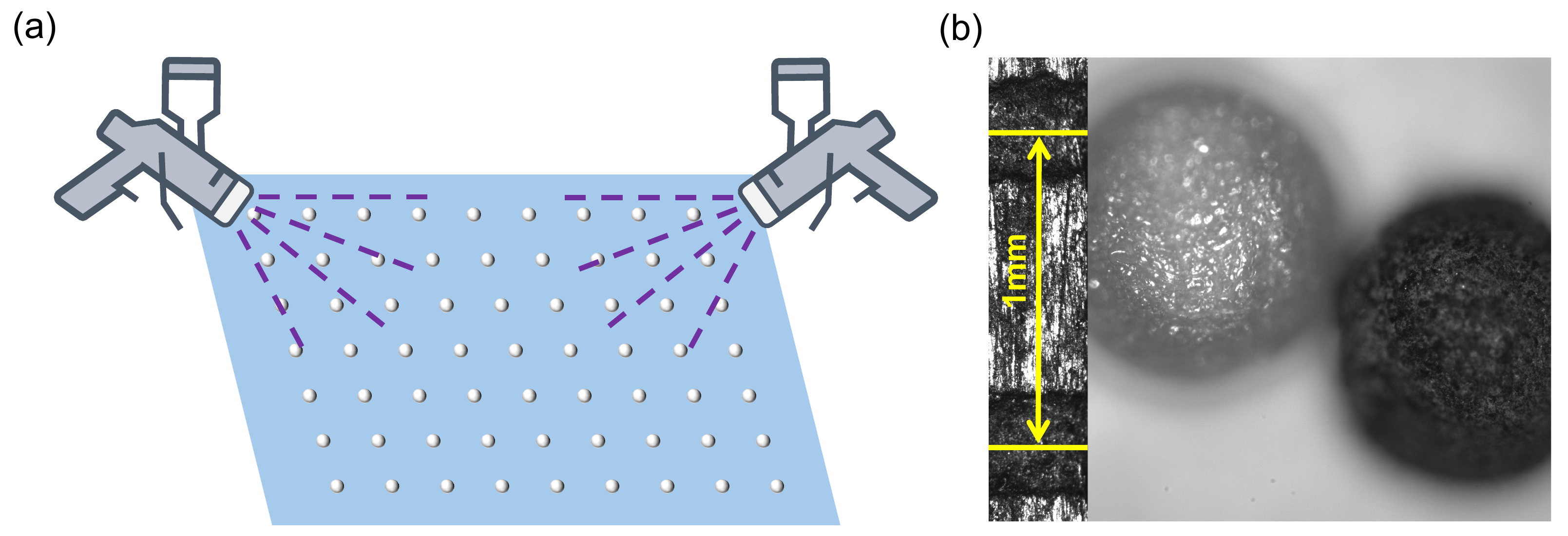} 
    \caption{(a) Schematic illustration of the fabrication process for magnetic particles. Polystyrene particles are secured on a flat plastic board using adhesive tape. Magnetic paint is manually sprayed from the top left and top right at a $45^\circ$ angle, coating the particles with a thin layer of magnetic material that can be magnetized in the presence of a magnetic field but exhibits no residual magnetism in free space.  (b) Optical microscopic image for the comparison of polystyrene particles before (on the left with brighter surface) and after (on the right with darker surface) coating. The magnetic layer increases mass and rigidity of the particle, and introduces a non-uniform surface pattern, which is essential for rotation tracking.}
    \label{fig:PaintParticles}

\end{figure}

In this study, we employ a handmade method to fabricate magnetic particles. Polystyrene (Styrofoam) particles, with a density of only 50 kg/m\textsuperscript{3}, are selected as the core material due to their lightweight nature. These core particles have diameters ranging from 0.5 mm to 1 mm. To impart magnetic properties, the polystyrene cores are coated with a thin layer of magnetic paint containing iron dust, which adheres to the particle surfaces through manual spraying. The fabrication process (Fig. \ref{fig:PaintParticles}a) begins by placing the polystyrene particles onto adhesive tape affixed to a flat plastic board. Manual spraying from two sides at a $ 45^\circ $ angle relative to the board not only magnetizes the particles but also creates a distinguishable surface pattern due to the inherent non-uniform distribution of magnetic powder in the paint. Figure \ref{fig:PaintParticles}b presents an optical microscope image captured using an Olympus U-CA magnification changer, comparing the polystyrene particles before and after coating.

The particles are then sorted using a series of fine sieves with progressively smaller mesh sizes. Particles retained between the 0.696 mm and 0.828 mm sieves are selected for our experiments. The mean diameter is estimated to be 0.762 mm, calculated as the average of the mesh sizes of the two sieves. Batch weighing of the particles indicates an average density of $ 208 \pm 14 $ kg/m\textsuperscript{3}. While the magnetic coating layer significantly increases the overall particle density, it remains only about one-fifth of the density of water. Moreover, the magnetic paint does not uniformly coat the entire surface of each polystyrene particle due to several factors: the bottom side adhering to the tape during spraying, variations in the mixing and dilution of the magnetic paint, and minor inconsistencies in the spraying process. These factors, combined with the imperfect spherical shape of the particles, introduce magnetic anisotropy. This anisotropy allows the particles to follow the rotation of the external planar rotating magnetic field generated in the setup, as the magnetic moment of each particle naturally tends to align with the orientation of the field \cite{Shanko2022}. The driving magnetic torque is given by  

\begin{equation}
     \bm{T_\text{m}} = \bm{m_\text{p}} \times \bm{B},
\end{equation}
where $\bm{m_\text{p}}$ represents the magnetic moment of the particle induced by $\bm{B}$. Due to the magnetic anisotropy of the particle,  $\bm{m_\text{p}}$ is always slightly misaligned with $\bm{B}$, resulting in a nonzero magnetic torque that drives the particle to rotate following the rotating magnetic field. The magnetic anisotropy can be quantified as $ \Delta \chi = \chi_{\parallel} - \chi_{\perp} $, where $ \chi_{\parallel} $ denotes the magnetic susceptibility along the direction in which the magnetic field induces the strongest magnetic moment, and $ \chi_{\perp} $ is the susceptibility in a direction perpendicular to it \cite{Tierno2009}. In this study, we assume axisymmetric magnetic anisotropy, same as that exhibited by an ellipsoidal particle, as it has been shown in previous studies that this assumption can accurately represent the magnetic behavior of many simple particle geometries \cite{Beleggia2006}, meaning that $ \chi_{\perp} $ is uniform in all directions orthogonal to $ \chi_{\parallel} $. Then the magnitude of the magnetic torque can be written as \cite{Abbott2007}

\begin{equation}
     |\bm{T_m}|_{max} = \frac{1}{\mu_0} V_{\text{p}} \Delta\chi|\bm{B}|^2.
     \label{eq:Magnetic torque chi}
\end{equation}

Given the polydispersity of our magnetic particles in terms of size, shape, and magnetic properties, their magnetic susceptibility and anisotropy are statistically characterized using a Superconducting Quantum Interference Device (SQUID) magnetometer manufactured by Quantum Design Inc. The mean volume magnetic susceptibility $ \overline{\chi}_V $ is determined via batch measurements of 20 particles randomly oriented in the sample holder, yielding a mean value of $ |\overline{\chi}_V | = 0.136 \pm 0.046 $. Magnetic anisotropy is assessed through single-particle measurements, which give an average value of $ \Delta \chi = 0.011 \pm 0.002 $. The measurement procedure is further detailed in the appendix \ref{Appendix2}. Although $D_\text{p}$ varies within the range [0.696, 0.828] mm, the torque balance between magnetic and hydrodynamic contributions that is represented by the equivalence of \ref{eq:Stokes torque} and \ref{eq:Magnetic torque chi} is unaffected by $V_\text{p}$, since the $D_\text{p}^3$ terms cancel out with $V_\text{p} = \pi D_\text{p}^3 / 6$. In contrast, the variability of $\Delta\chi$ directly influences the magnitude of the magnetic torque $|T_\text{m}|$, but does not significantly impact the overall rotational dynamics in a statistical sense. This conclusion is supported by our experimental observations in a separate study across the explored parameter space, which consistently exhibit all three characteristic regimes of particle rotation: turbulence dominated, magnetism dominated, and an intermediate regime where both effects are comparable \cite{Wang2025}.

\section{Image processing and particle rotation tracking} \label{sec:Code}
Our algorithm, which determines all three components of the angular velocity of magnetic particles, is based on a 3D rotation tracking code that computes the rotation of spherical particles using only 2D images \cite{Niggel2023}. It achieves this by projecting the circular surface pattern of a particle from an image onto a hemispherical surface and then identifying the optimal angle through cross-correlation, aligning the pattern with that of the subsequent instant. Therefore, only one camera is required in the experiments. 

\begin{figure}[htbp]
    \centering
    \includegraphics[width=0.6\textwidth, angle=0]{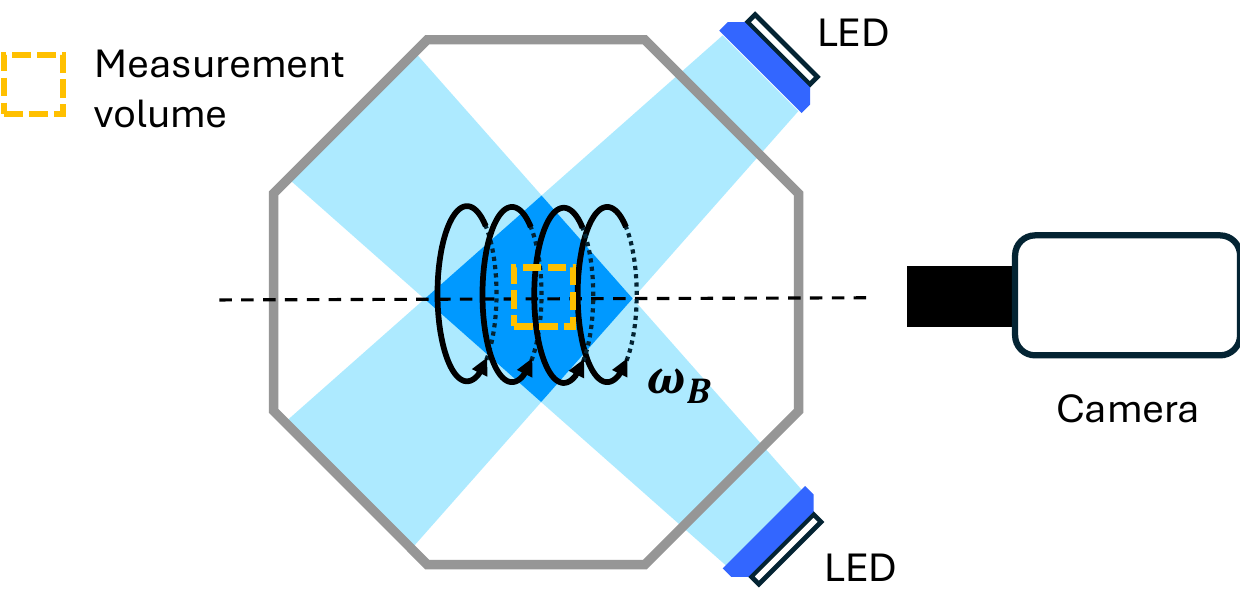} 
    \caption{Top view of the one-camera layout for 3D rotation tracking. The optical axis of the camera is aligned with the direction of the angular velocity $\bm{\omega_B}$ of the rotating magnetic field, as indicated by the black dashed line. This alignment ensures that the magnetic field influences only the particle angular velocity along this direction, while turbulence, if present, affects all three components. The orange dashed square marks the measurement volume, which has a side length of 12 mm. Two blue LEDs illuminate the measurement volume symmetrically from both sides, each positioned at an angle of $45\degree$ relative to $\bm{\omega_B}$.}
    \label{fig:One camera lauout}
\end{figure}

Figure \ref{fig:One camera lauout} illustrates the one-camera optical system used for tracking the rotation of magnetic particles. The system employs a Photron NOVA S6 high-speed camera, positioned along the rotation axis of the planar rotating magnetic field with angular velocity $\bm{\omega_B}$, and records at a frame rate of 3000 fps with a resolution of 1024$\times$1024 pixels. The alignment of the camera optical axis with the rotation axis of the magnetic field ensures that the magnetic interaction influences only this component of the particle angular velocity, whereas turbulence affects all three components. The camera is aligned perpendicularly to one of the observation windows of the octagonal prism water tank. This orientation minimizes the effects of refraction that occur as light passes from water to air through the walls of the water tank. To clearly visualize the surface patterns on the magnetic particles that is essential for accurate rotation tracking, two blue LEDs with a wavelength of 495 nm are used for illumination. These LEDs are symmetrically placed on either side of the measurement volume, each set at an angle of $45\degree$ relative to the optical axis of the camera. The side length of the measurement volume is approximately 12 mm.

\begin{figure}[htbp]
    \centering
    \includegraphics[width=0.9\textwidth, angle=0]{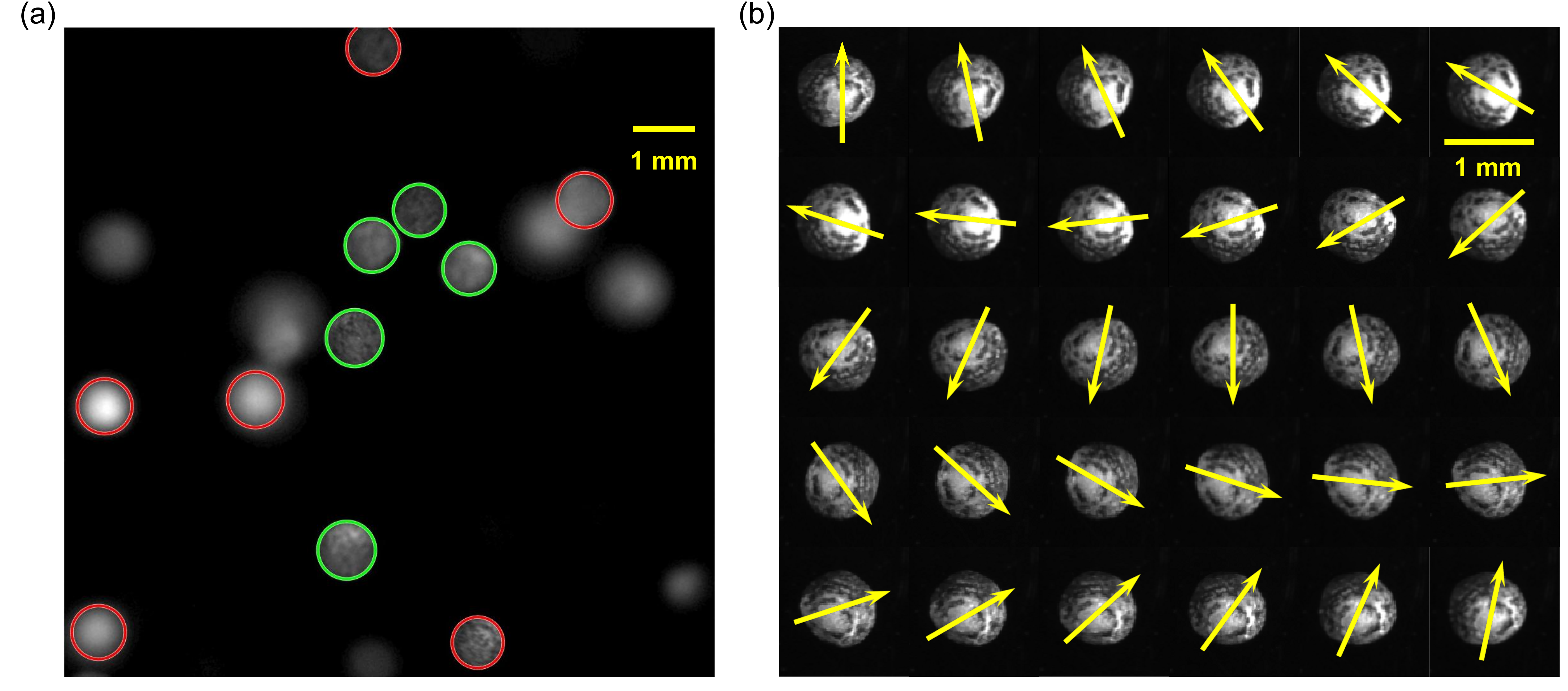} 
    \caption{(a) An original image showing particles with radii within the acceptable range, marked by red and green circles. The autofocusing algorithm identifies well-focused particles, which are highlighted in green. (b) An example of a well-focused rotating particle. This particle appears as an 80-pixel-diameter object in the images and rotates counterclockwise within the image plane. The sequence of single-particle frames progresses from left to right, top to bottom, with the yellow arrow in each frame indicating the particle orientation.}
    \label{fig:ParticleIllustration}
\end{figure}

In the original images, the particles illuminated and captured by the camera are not always sharply in focus due to the limited depth of field imposed by optical constraints. The initial step involves identifying particles that exhibit a well-defined circular shape and a diameter within the specified range of 0.8–1.2 $D_\text{p}$. Particles that are severely blurry typically fail to meet this criterion as their apparent diameters deviate significantly from the expected size due to their axial displacement relative to the focal plane. In Fig. \ref{fig:ParticleIllustration}a, particles that have passed this filter are highlighted with red and green circles. A 2D particle tracking algorithm \cite{Crocker1996}, followed by a particle cropping procedure, is then applied based on the identified positions to extract single-particle images across consecutive frames. Since many out-of-focus particles still remain, a focus detection step is essential to further filter out blurry particles (marked with red circles in Fig. \ref{fig:ParticleIllustration}a), ensuring that only high-quality single-particle images are fed into the particle rotation tracking algorithm. This autofocusing detection method is discussed in more details below.

The 3D rotation tracking code is first validated in a quiescent flow environment under the rotating planar magnetic field in the setup, with frequencies $f_\text{m}$ ranging from 5 to 45 Hz in 5 Hz increments. Magnetic particles are placed in a separate transparent cylindrical container filled with water. The container has a diameter of 80 mm, a height of 40 mm, and negligible wall thickness, and is positioned at the center of the otherwise empty water tank. It is sealed with a transparent lid. Due to their lower density relative to water, the particles float upward and distribute evenly beneath the lid, forming a single layer without overlapping or contacting one another. To ensure that the camera captures the particle rotation along the axis of the magnetic field, the entire water tank and coils are rotated by 90$\degree$, allowing the camera to film from above through the lid of the cylindrical container. This small cylindrical container confines the magnetic particles within the designated measurement volume of the setup. Therefore, the rotation of particles is driven solely by the magnetic field, with the magnetic torque acting to rotate the particles against the hydrodynamic torque from the surrounding quiescent water. In this controlled condition, once the particles begin rotating continuously, their rotation frequency must equal the imposed magnetic field frequency $f_\text{m}$.

\begin{figure}[htbp]
    \centering
    \includegraphics[width=0.8\textwidth, angle=0]{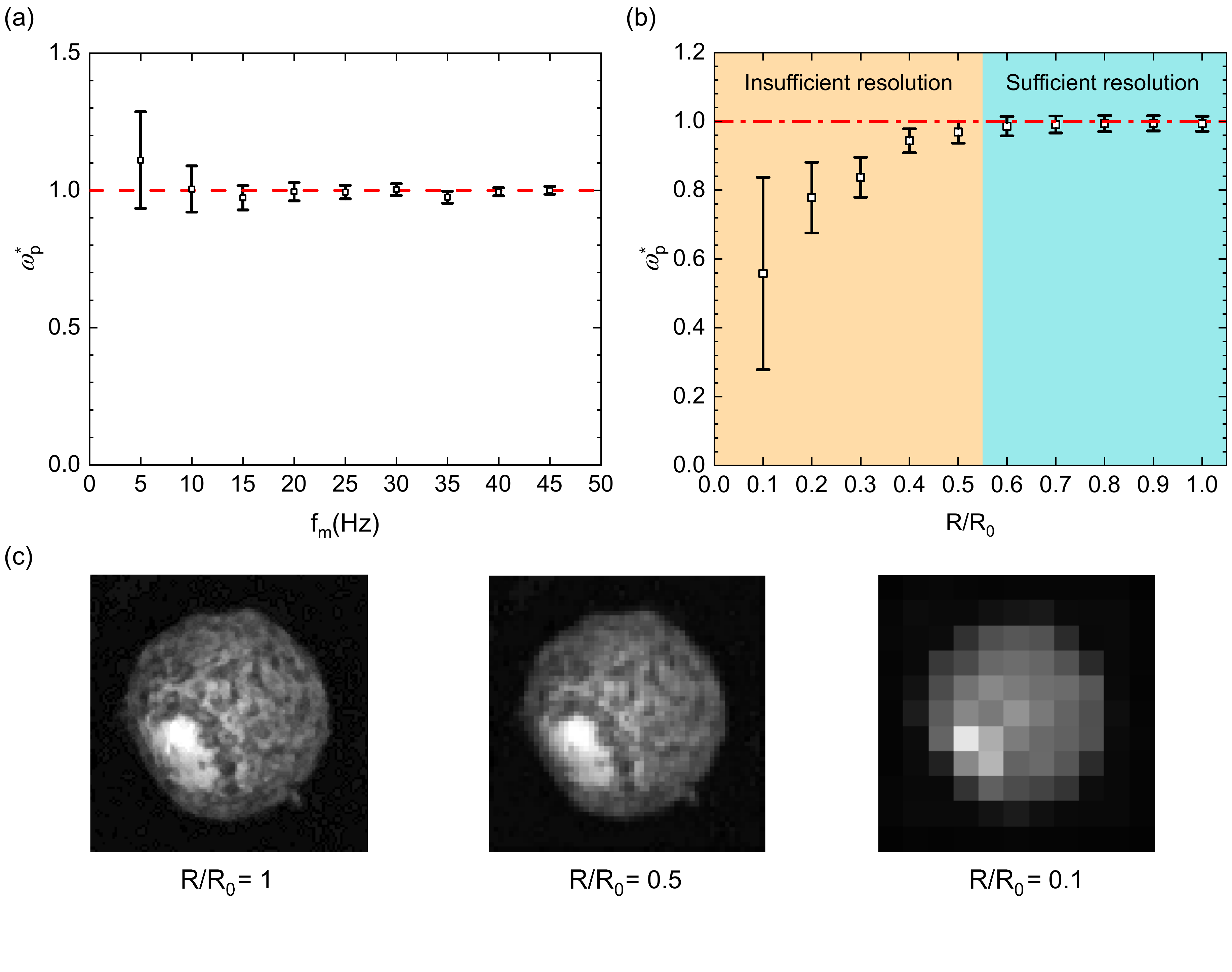} 
    \caption{Rotation tracking validation. The red dashed line in each figure represents $\omega^*_\text{p} = 1$. Square data points indicate the mean values of $\omega^*_\text{p}$ for each case, and error bars represent $\pm 1$ standard deviation. (a) Normalized rotation frequencies $\omega^*_\text{p}$ of particles computed by the tracking code at various rotating magnetic field frequencies. (b) $\omega^*_\text{p}$ as a function of the ratio $R / R_0$, where $R$ is the one-dimensional image resolution and $R_0 = 110$ pixels corresponds to the resolution used in the experiments. All data correspond to particles rotating at 20 Hz. Images with a normalized resolution ratio $R/R_0 \geq 0.6$ offer sufficient detail for accurate rotation tracking. (c) Sample images at three different resolutions. The image with $R/R_0 = 1$ clearly displays the surface pattern, while the one with $R/R_0 = 0.5$ still reveals the pattern but with noticeable edge blurring. In contrast, the image with a resolution of 11$\times$11 pixels presents only a single bright pixel at the location of the surface pattern, indicating a complete loss of trackable features.}
    \label{fig:Validation_resolution}
\end{figure}

Figure \ref{fig:Validation_resolution}a presents the tracking results of the magnitude of the normalized angular velocity $\omega^*_\text{p} = {\omega}_\text{p} / (2\pi f_\text{m})$ along the direction of the magnetic field. Therefore, $\omega^*_\text{p} = 1$ indicates that the rotation of particles is synchronized with that of the magnetic field. Each data point represents an average of approximately 5000 rotation measurements, evenly contributed by 20 particles. The results clearly demonstrate that the rotation tracking code achieves high accuracy when applied to our magnetic particles across the nine tested frequencies. The maximum error $|\overline{\omega^*_\text{p}} - 1|$ is 0.11 at $f_\text{m} = 5$ Hz and as $f_\text{m}$ increases, this tracking error drops rapidly to $5 \times10^{-5}$ at $f_\text{m} = 45$ Hz.

All rotation tracking results discussed in this paper are computed from recordings with an original resolution of 105 pixels/mm. For processing convenience, each single-particle image is cropped to a size of 110$\times$110 pixels. To evaluate the effect of image resolution on rotation tracking accuracy, we tested particles rotating with a 20 Hz magnetic field using images of varying resolutions. Lower-resolution images are generated from the original recordings using a downsampling method with linear interpolation. As shown in Fig. \ref{fig:Validation_resolution}b, the rotation tracking code remains accurate with images reduced to a resolution of 66$\times$66 pixels, yielding $\omega^*_\text{p} = 0.99 \pm 0.03$. Since the ideal value $\omega^*_\text{p} = 1$ lies within the error bounds, such images are considered to provide sufficient resolution for reliable rotation tracking. Although images with a resolution of 55$\times$55 pixels appear noticeably blurry, they still retain a discernible surface pattern. However, the result drops slightly to $\omega^*_\text{p} = 0.97 \pm 0.03$, with the upper bound of the error margin just reaching $\omega^*_\text{p} = 1$. Therefore, these images, along with those at even lower resolutions that yield $\omega^*_\text{p}$ values further deviating from 1, are not considered suitable for accurate rotation tracking.

\begin{figure}[H]
    \centering
    \includegraphics[width=1\textwidth, angle=0]{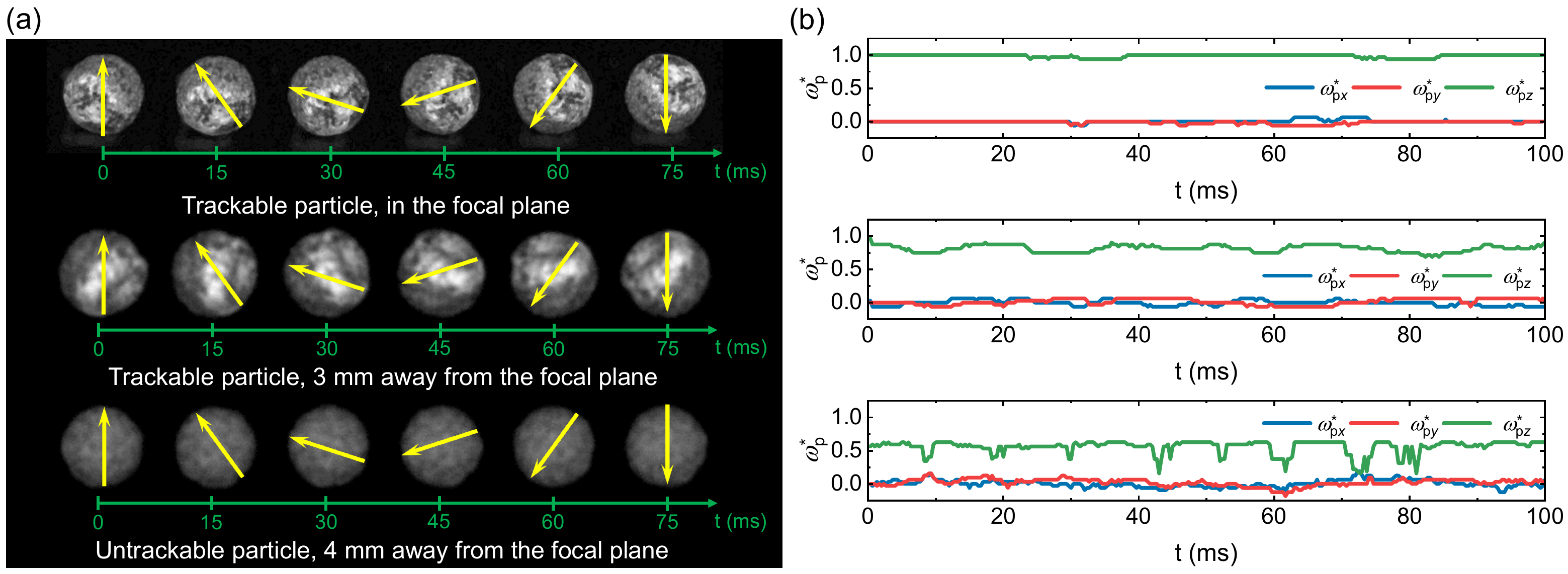} 
    \caption{(a) Examples of the three types of particle images. Yellow arrows indicate the orientation of the particle in each frame. For each particle, six frames over a duration of 75 ms are presented. (b) The corresponding tracking results of the three example particles shown in (a), presenting the normalized angular velocity $ \omega^*_\text{p} $ at $ f\text{m} = 20 $ Hz. The first row (top) shows a well-focused particle recorded in the focal plane, exhibiting an almost constant $ \omega^*_\text{p} $ with a mean value of 0.99 and a standard deviation of 0.01. The second row (middle) displays a particle recorded 3 mm away from the focal plane, with a mean $ \omega^*_\text{p} $ of 0.81 and a standard deviation of 0.05. Its tracking results indicate clear rotational behavior, and it is therefore classified as a trackable particle. The third row (bottom) presents a heavily blurry particle recorded 4 mm away from the focal plane, with a mean $ \omega^*_\text{p} $ of 0.56 and a standard deviation of 0.10. Due to the large deviation and fluctuation in its tracking results, this particle is classified as rotationally untrackable.}
    \label{fig:Trackable particles}
\end{figure}

The validation results shown in Figure \ref{fig:Validation_resolution} demonstrate that, despite the polydispersity in particle size ($D_\text{p}$) and variability in magnetic anisotropy ($\Delta\chi$), the rotation tracking accuracy remains sufficient as long as the images are well-focused and of adequate resolution. To objectively assess image focusing and ensure that only high quality single particle images are used for rotation tracking, a final autofocusing detection step is incorporated into the image preprocessing pipeline. Although the rotation of well-focused particles in the focal plane, captured with sufficient resolution, can be accurately tracked with nearly constant angular velocity and small standard deviations, some blurry particles outside the focal plane also exhibit nearly constant angular velocities, albeit with slightly lower mean values. These particles are considered rotationally trackable, though with reduced accuracy. In contrast, other blurry particles are classified as rotationally untrackable because their tracking results show angular velocity mean values that deviate significantly from the true rotation speed and exhibit large standard deviations. To assess this more quantitatively, we define a selection criterion to identify untrackable particles based on the angular velocity along the \textit{z} axis: if the mean deviates by more than 0.2 from the expected $ \omega^*_\text{p} = 1$, indicating a relative error of 20\%, or, alternatively, if the standard deviation exceeds 0.1. Figure \ref{fig:Trackable particles} presents an example of each of these three types of particles. All particles follow a rotating magnetic field along the \textit{z} axis, which is perpendicular to and pointing out of the image plane. 

\begin{figure}[H]
    \centering
    \includegraphics[width=1\textwidth, angle=0]{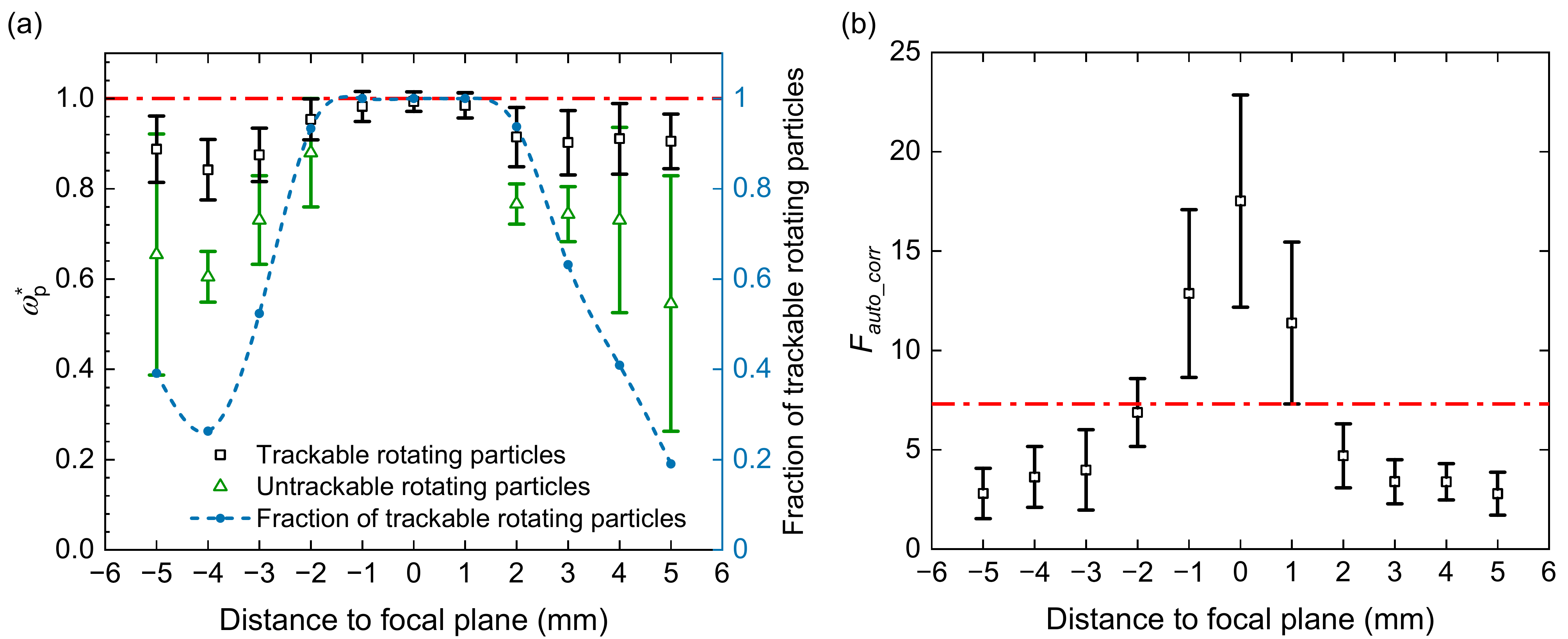} 
    \caption{(a) The mean and spread of normalized angular velocities $\omega^*_\text{p}$ for trackable particles (including both in-focus and out-of-focus) and untrackable particles at various imaging planes. Blue circles connected by dashed lines represent the ratio of trackable particles to the total number of recorded particles at each plane. The red dot-dashed reference line indicates $\omega^*_\text{p} = 1$ and a trackable particle fraction of 1. (b) The mean and spread of image definition, evaluated using the $ F_{\text{auto\_corr}} $ metric, for particles at each imaging plane. The red dot-dashed reference line represents the threshold value of $ F_{\text{auto\_corr}} = 7.3$.}
    \label{fig:Trackable_autocor}
\end{figure}

To further quantify the relationship between image focus and the trackability of particles, a comparative experiment is conducted by placing particles at different axial positions relative to the focal plane. These positions lie along the depth of field, which is aligned with the rotation axis. Figure \ref{fig:Trackable_autocor}a presents the tracking results of rotationally trackable and untrackable particles rotating under a magnetic field at $f_\text{m} = 20$ Hz, plotted as a function of their distance from the focal plane. The fraction of trackable particles relative to the total number of observed particles remains close to 1 when the distance from the focal plane is within 1 mm, but drops sharply to 0.4 when the distance exceeds 4 mm. This indicates that particles located within 1 mm from the focal plane provide sufficient accuracy for reliable rotation tracking. These particles are consequently classified as in-focus particles. In contrast, trackable particles in other imaging planes show mean values of $\omega^*_\text{p}$ that deviate significantly from 1. Therefore, these particles, regardless of whether they are trackable or untrackable, are considered out of focus because their images do not provide reliable tracking accuracy. To distinguish in-focus particles from out-of-focus ones, it is necessary to use a metric that quantitatively evaluates image definition. For this purpose, we adopt a sharpness metric based on the auto-correlation of pixel gray levels, as proposed by Vollath \cite{Vollath1988}.

\begin{equation}
    F_{\text{auto\_corr}} = \sum_y^Y \sum_x^{X-2} i(x,y) \cdot i(x+1,y)
    - \sum_y^Y \sum_x^{X-2} i(x,y) \cdot i(x+2,y),
\end{equation}
where $Y$ and $X$ are the horizontal and vertical sizes of the image, respectively, and $ i(x,y) $ denotes the gray level intensity of pixel $ (x,y) $. This metric calculates the difference between the auto-correlations of each pixel and its first and second neighboring pixels along the \textit{x} axis. Therefore, images with sharp, high-frequency textures produce higher $ F_{\text{auto\_corr}} $ values. Given that the single-particle image resolution is 110$\times$110 pixels, the range of $x$ is set from 1 to 108 to ensure that the computation stays within the image bounds.

The $ F_{\text{auto\_corr}} $ values for all particles at different imaging planes, as measured in Fig. \ref{fig:Trackable_autocor}a, are shown in Fig. \ref{fig:Trackable_autocor}b. The particles close to the focal plane exhibit significantly higher values for $ F_{\text{auto\_corr}} $ than those further away. The wide spread in these values is attributed to the diversity of particle surface patterns and the constantly changing orientations during rotation, since $ F_{\text{auto\_corr}} $ only considers pixel correlations along the \textit{x} axis. Because the images have a black background where $i(x, y) = 0$ and a bright foreground (the particle) where $i(x, y) > 0$, the value of $F_{\text{auto\_corr}}$ is always positive even when the particle is completely out of focus. In such cases, the intensity in the foreground becomes a uniform positive constant. Taking the focal plane and the two imaging planes closest to it as reference, where no untrackable particles are observed, a threshold of $F_{\text{auto\_corr}} = 7.3$ is applied in our experiments. This value corresponds to the lower limit of the $F_{\text{auto\_corr}}$ measured for particles in these three planes and is used to identify in-focus particles suitable for reliable rotation tracking.

\section{Proof of principle: rotation statistics of magnetically actuated particles in turbulence} \label{sec:experiments}

We investigate the rotational dynamics of magnetic particles actuated by an external rotating magnetic field while immersed in turbulence through experiments using the method introduced in the previous sections, as well as through numerical simulations \cite{Wang2025}. In this section, we provide a consistent comparison between experimental and numerical results, which serves to validate both the experimental setup and the methodology employed in this study.

\begin{figure}[htbp]
    \centering
    \includegraphics[width=0.7\textwidth, angle=0]{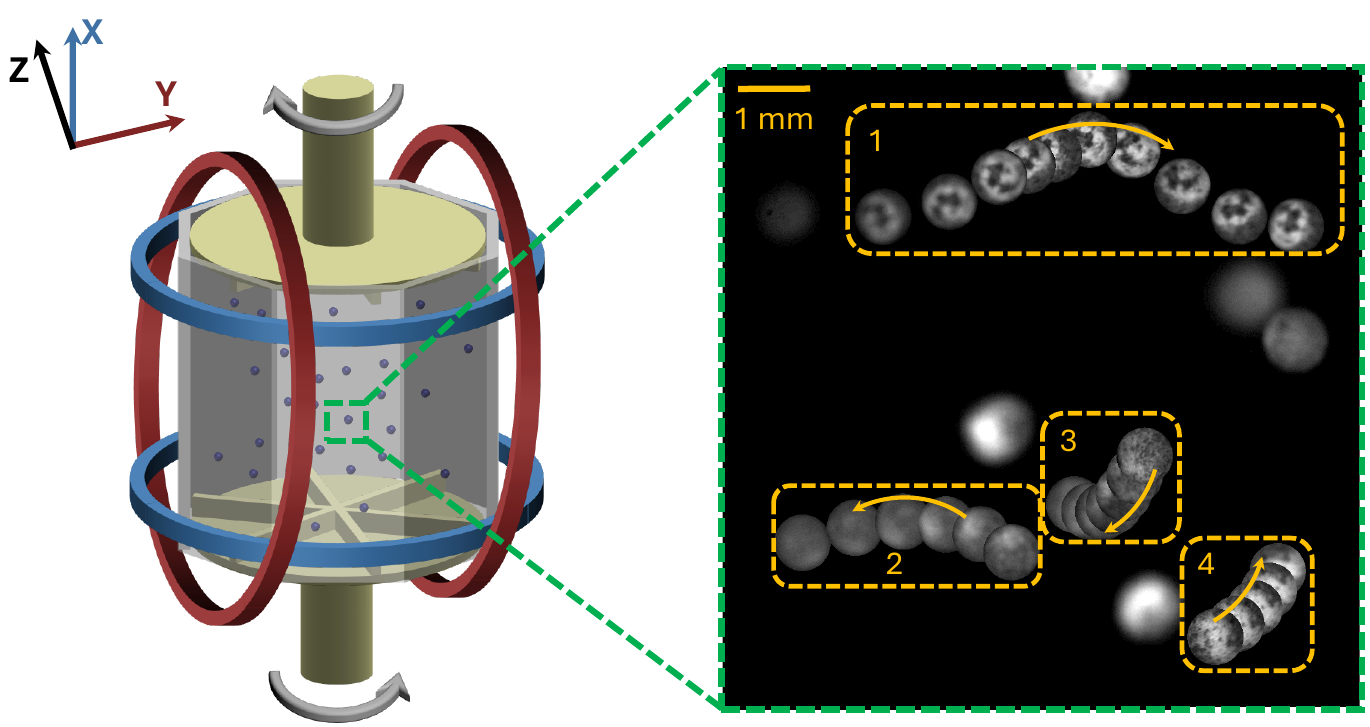} 
    \caption{On the left is a schematic illustration of the experimental setup, consisting of a French washing machine that generates turbulent flow and two pairs of Helmholtz coils aligned along the \textit{x} and \textit{y} axes to produce a rotating planar magnetic field in the \textit{xy} plane. The measurement volume is indicated by the green square. To the right, a sample image shows ten particles within a field of view measuring 12 mm × 12 mm, but only 4 particles pass the circle detection. A time-lapse visualization displays the same region over 200 consecutive frames. For clarity, only one out of every twenty frames is displayed, with the 2D trajectories of four particles overlaid. An arrow in each dashed-line rectangle indicates the motion direction of the corresponding particle. The number of frames in which each particle appears varies due to differences in the duration they remain in focus. This tracking step enables the identification of individual particles across frames, thereby facilitating the extraction of single-particle image sequences for subsequent rotational analysis.}
    \label{fig:Experiment}
\end{figure}

The experimental investigation of the rotational motion of magnetic particles is conducted with both impellers rotating at 0.83 Hz, generating a turbulent flow with a Kolmogorov length scale of $ \eta = 0.072 $ mm, and time scale of $ \tau_{\eta} = 5.20 $ ms, as discussed in Sec. \ref{sec:Integrated design}. Magnetic particles are introduced with a volume fraction of $\Phi_V =7 \times 10^{-4}$ in the measurement volume, and recordings are taken within the central measurement volume of the setup. As illustrated in Fig. \ref{fig:Experiment}, the particles are imaged from the perspective of the optical axis. 2D tracking is applied to identify individual particles and reconstruct their 2D trajectories, enabling the extraction of single-particle images from each trajectory, which are then used as input for the rotation tracking algorithm. In this experiment, a 15-second recording at 3000 fps yields approximately 5000 measurements from an average of 170 particles. This implies that the average in-focus duration of a particle within the measurement volume is only about 10 ms. For the purpose of studying the rotational dynamics of particles, we note that our particle volume fraction, $\Phi_V = 7 \times 10^{-4}$, lies near the threshold between the one-way and two-way coupling regimes \cite{Elghobashi1994, Baker2019, Crowe2011, Poelma2007}. Therefore, we do not expect significant changes in the influence of turbulence on the particles' rotational dynamics regardless of whether the magnetic field is applied or not.

\begin{figure}
    \centering
    \includegraphics[width=0.5\textwidth, angle=0]{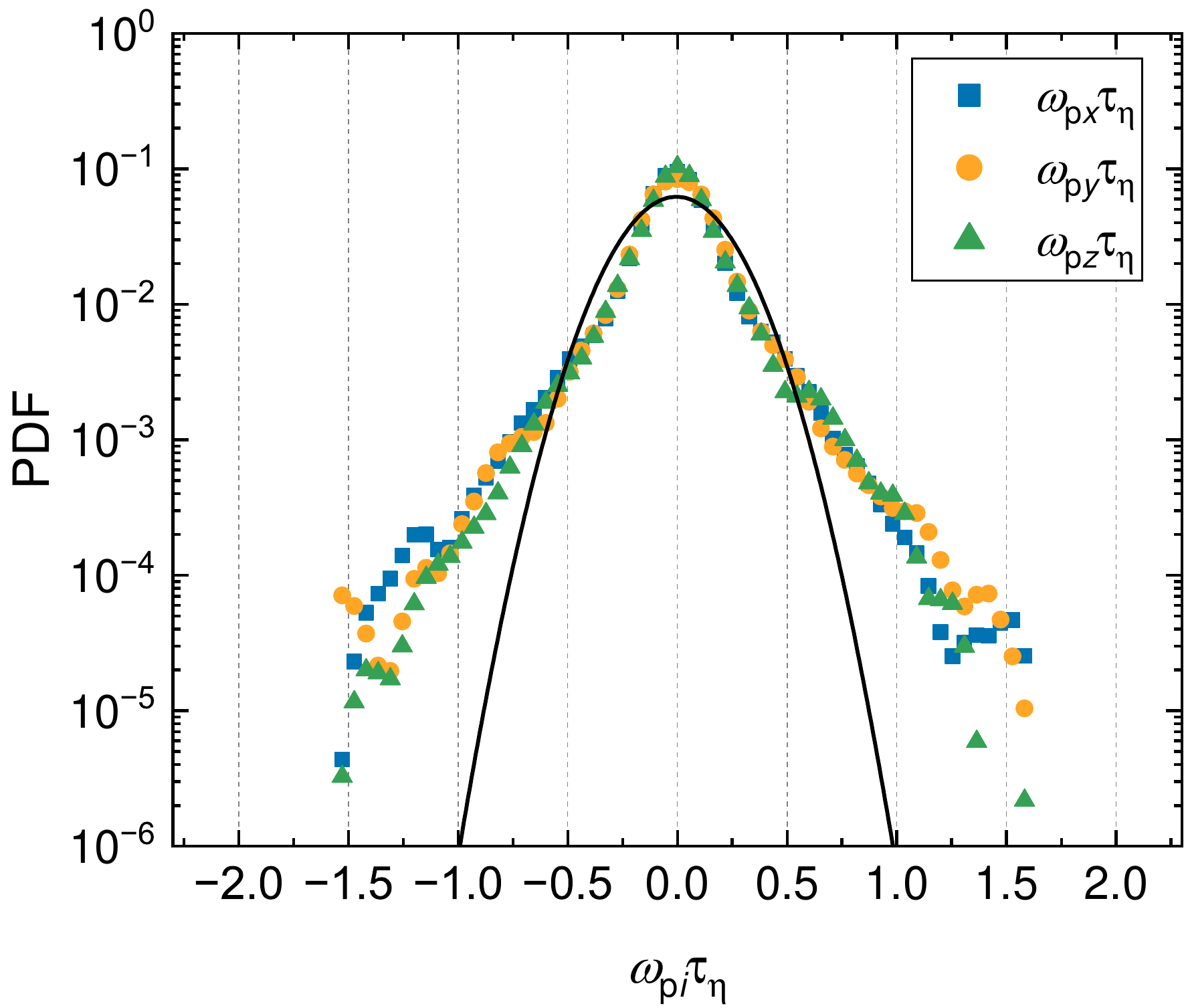} 
    \caption{Probability distribution functions (PDF) of the angular velocities of particles $\omega_{\text{p}i}$ $(i = x, y, z)$ components in turbulence without the presence of a magnetic field. The results comprise 150,000 measurements contributed by approximately 5,000 particles. The black curve represents a Gaussian distribution with the same mean and variance as $ \omega_{\text{p}x} $, included for comparison.}
    \label{fig:PDFs without magnetic field}
\end{figure}

In the absence of a rotating magnetic field, the angular velocities of the particles $\omega_{\text{p}i}$ $ (i = x, y, z)$ are resolved with a resolution of 10.5 rad/s over the range [–314.2, 314.2] rad/s, corresponding to $f_\text{p} = \omega_\text{p}/2\pi \in [-50, 50]$ Hz. The resolution and range are constrained by the available computational memory, with this configuration occupying approximately 40 GB. As shown in Fig. \ref{fig:PDFs without magnetic field}, the probability density functions (PDFs) of the three components $ \omega_{\text{p}i} $ normalized by $\tau_\eta$ are nearly identical and symmetric, yet display heavier tails compared to a Gaussian distribution with the same mean and variance. This deviation arises because the particle angular velocity is directly related to the local vorticity in turbulence,which is known to exhibit strong intermittency at length scales within the dissipation range \cite{Sreenivasan1997}.

\begin{figure}
    \centering
    \includegraphics[width=1\textwidth, angle=0]{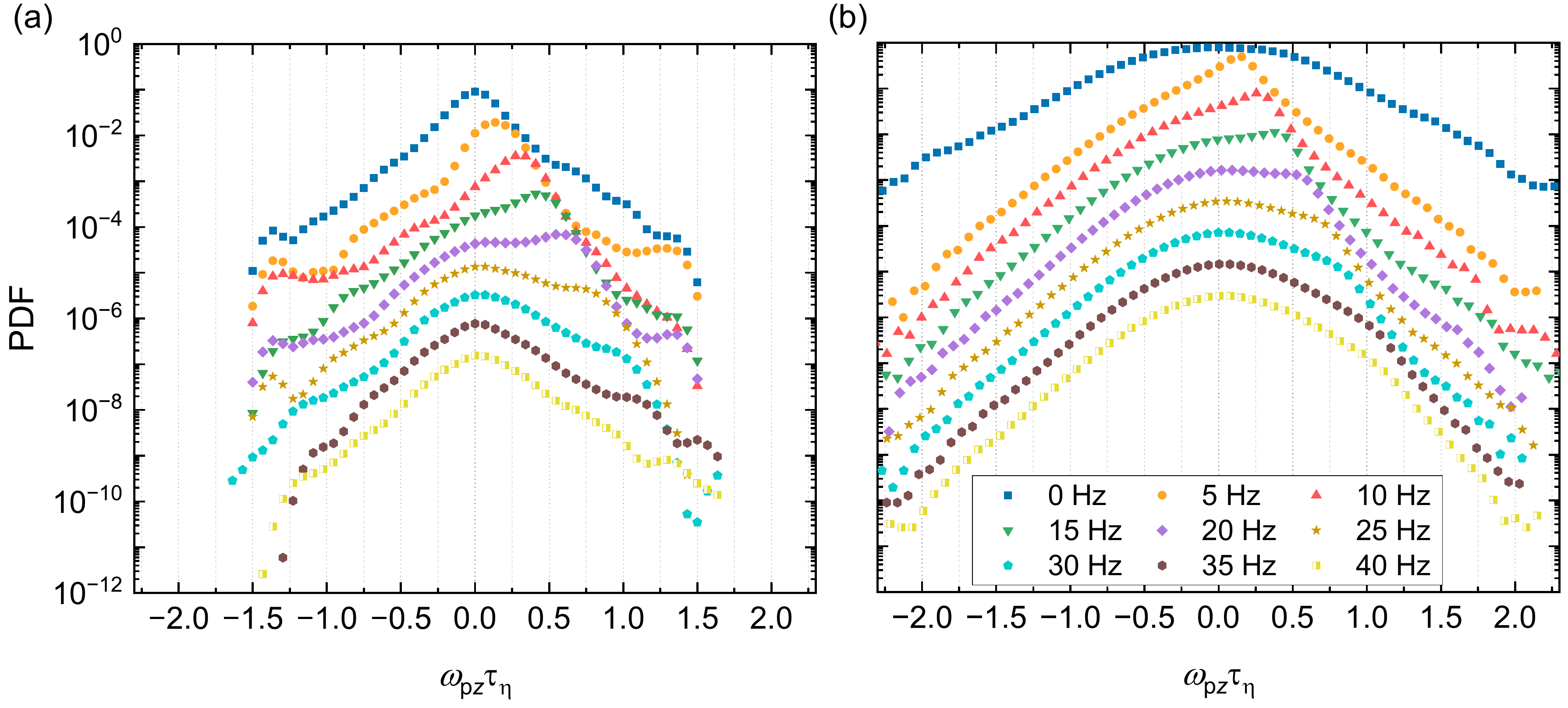} 
    \caption{Experimental (left) and numerical (right) results of rotation tracking of light magnetic particles under varying magnetic field frequencies $f_\text{m}$. (a) Experimental results showing the rotation frequency of magnetic particles along the $ z $ axis, $ \omega_{\text{p}z} $ normalized by the Kolmogorov time scale $ \tau_\eta = 5.21 $ ms. The particles are subjected to rotating magnetic fields with frequencies $f_\text{m}$ ranging from 5 Hz to 40 Hz in 5 Hz increments, all with a constant magnetic flux density of 1.6 mT. The 0 Hz case represents pure turbulence without magnetic forcing. (b) Numerical results of particle rotation are obtained under the same sequence of magnetic field frequencies and magnetic flux density as in the experiments, but within a weaker turbulent flow with a Kolmogorov time scale of $ \tau_\eta = 100 $ ms \cite{Wang2025}. PDFs in (a) and (b) are plotted with a vertical shift for clarity.}
    \label{fig:pdfs}
\end{figure}

When both turbulence and a rotating magnetic field with $|\bm{B}| = 1.6$ mT are present, the rotational dynamics of magnetic particles are influenced by both effects: the magnetic field tends to drive the particles to rotate at its own frequency $ f_\text{m} $, while turbulence introduces stochastic perturbations to the rotation. As a result, the rotating planar magnetic field induces a distinct peak in the PDF of $ \omega_{\text{p}i} $ at $ \omega_{\text{p}z} = 2\pi f_\text{m} $. Figure ~\ref{fig:pdfs}a presents the experimental results, showing that the magnetism-induced peak consistently appears at each tested $f_\text{m}$, but its amplitude decreases as $f_\text{m}$ increases. Meanwhile, the turbulence-induced peak remains centered at zero and becomes increasingly prominent, eventually surpassing the magnetism-induced peak. This indicates that for most particles and during most of their residence time in the measurement volume, rotational dynamics are primarily governed by turbulence rather than magnetic forcing. The transition in dominance from magnetic to turbulent control occurs somewhere in the range $20 < f_\text{m} < 25$ Hz and can be attributed to two main factors. Primarily, at higher $f_\text{m}$, particles have less time to accumulate the angular momentum needed to match the rotation of the magnetic field. This provides turbulence with a greater opportunity to interfere and dominate their rotational dynamics. In addition, the polydispersity in the anisotropic magnetic susceptibility $\Delta \chi$ among particles results in fewer particles possessing sufficient $\Delta \chi$ to generate the magnetic torque $\bm{T}_\text{m}$ necessary to follow faster magnetic field rotations.

To gain a clearer understanding of the mechanisms underlying this dynamical system, numerical simulations are conducted to resolve the overdamped rotational motion of magnetic particles, governed by the balance $\bm{T}_\text{m} + \bm{T}_\text{St} = \bm{0}$. In the Stokes regime, the hydrodynamic drag torque $\bm{T}_\text{St}$ is also approximated using Eq. (\ref{eq:Stokes torque}), but with $\delta\omega$ replaced by $(\bm{\omega}_\text{p} - \bm{\omega}_\text{f})$. Here, $\bm{\omega}_\text{f}$ denotes half the local fluid vorticity evaluated at the particle’s center of mass $O$, that is, $\bm{\omega}_\text{f} = \frac{1}{2}[\bm{\nabla} \times \bm{u}_\text{f}]_O$. It is important to note that particle volume has no effect in this simulation, as both $\bm{T}_\text{m}$ and $\bm{T}_\text{St}$ are proportional to the particle volume $V_{\text{p}}$, which cancels out in the torque balance. The rotational motion is integrated using a third-order Runge–Kutta method, with $\bm{\omega}_\text{f}$ obtained from Direct Numerical Simulations (DNS) performed via a pseudo-spectral method. While the comparison to DNS provides useful insights, it should be noted that the DNS is performed at a significantly lower Reynolds number $\text{Re}_\lambda = 60$ compared to $\text{Re}_\lambda = 398$ in the experiments, which may limit the quantitative agreement with our experimental results. To account for particle polydispersity, the magnetic susceptibility anisotropy $\Delta\chi = \chi_\parallel - \chi_\bot$ is modeled as a Gaussian distribution with a mean of 0.01 and a standard deviation of 0.002, which are the same as the distribution of $\Delta \chi$ of the magnetic particles used in experiments. As shown in Fig. \ref{fig:pdfs}b, despite the smaller Taylor Reynolds number $\text{Re}_\lambda = 60$ in our DNS, which causes the PDFs to be more broadly distributed and exhibit weaker intermittency compared to the experimental results, the simulations successfully reproduce the same trend observed in experiments: a magnetism-induced peak appears and gradually decreases as $f_\text{m}$ increases. The transition in peak dominance from magnetic to turbulent control also occurs between $20 < f_\text{m} < 25$ Hz. This consistent transition point, despite differences in turbulence intensity, suggests that the magnetic anisotropy $ \Delta\chi$ plays a dominant role in governing the rotational dynamics of the particles in the given parameter space. Statistically, it is $\Delta \chi$ that governs the strength of magnetic interaction and thus determines whether the particles are primarily influenced by magnetic forcing or turbulent fluctuations. A comprehensive discussion of the particle rotational dynamics and the numerical methodology employed in this study is provided in our companion paper \cite{Wang2025}. 

\section{Conclusion and outlook}  \label{sec: Conclusion}
In this paper, we have presented an experimental method for tracking the rotational motion of particles that are significantly lighter than water, have sizes comparable to the dissipation scale, and exhibit anisotropic magnetism that enables them to follow the rotation of an external rotating magnetic field while remaining immersed in a turbulent flow.

The complete rotation tracking algorithm consists of several image processing steps, culminating in the application of a 3D rotation tracking code \cite{Niggel2023} that requires only 2D images captured by a single camera. This code resolves the angular velocities of particles in all three components by computing the optimal orientation between each pair of consecutive frames, based on the unique and distinguishable surface patterns present on the particles. Specifically, objects that retain a circular shape within a defined size range are first identified from the original images. Next, 2D particle tracking \cite{Crocker1996} across all consecutive frames, followed by cropping, yields single-particle image sequences suitable for rotation tracking. However, prior to applying the 3D tracking code, an autofocusing detection step is performed to ensure that only well-focused particles that exhibit clearly defined surface patterns are used for analysis.

The accuracy of this algorithm applied to our magnetic particles has been validated under varying conditions, including different magnetic field rotation frequencies $f_\text{m}$, different image resolutions $R$, and varying degrees of image blurriness, the latter is quantified using the $F_{auto\_corr}$ metric. Results demonstrate that for images with resolution $R = 110$ pixels and magnetic field frequencies in the range $f_\text{m} \in [5, 50]$ Hz, the maximum tracking error $|\overline{\omega^*_p} - 1|$ is less than 0.11. Even at lower resolutions down to $R = 66$ pixels, the method maintains sufficient accuracy, yielding results of $\omega^*_\text{p} = 0.99 \pm 0.03$. Images with $F_{auto\_corr} > 7.3$, corresponding to particles located within ±1 mm of the focal plane, also provide sufficiently high image quality, with a maximum observed error of $|\overline{\omega^*
_p} - 1| = 0.02$ in the tracking results.

Finally, experiments have been conducted in which the rotation of particles is influenced by both a turbulent flow and a rotating magnetic field at various frequencies $f_\text{m}$. The experimental results show good agreement with numerical simulations, both of which reveal that the rotational dynamics of magnetic particles exhibit a clear two-regime behavior: particles are predominantly governed either by turbulence or by the rotating magnetic field. The transition in dominance between these two mechanisms underscores the rich and complex dynamics that emerge from the competition between magnetic forcing and turbulent disturbances.

The rotation tracking method developed in this study provides a powerful tool for the detailed investigation of particle rotational dynamics in turbulence. Although a single camera provides sufficient information to resolve all three components of a particle angular velocity, adding more cameras will be beneficial. Multiple cameras can capture more particles within the measurement volume from different perspectives, and combining views of the same particle from different angles can further enhance tracking accuracy. Therefore, this approach can be applied to other turbulent flow configurations, such as Taylor–Couette flow, where tracking particles is highly constrained due to the limited optical access between the inner and outer walls. It may, for example, be used to study how non-colloidal suspended particles modulate flow transitions or enhance friction on the inner cylinder \cite{Kang2021, Kang2023}. In combination with our light magnetic particles, this method enables the study of complex rotational behavior governed simultaneously by the external rotating magnetic field and the turbulent flow. Furthermore, these particles serve as a platform for modulating turbulence from inside small scale vortex filaments via magnetic forcing. For instance, they can be used to explore the attenuation or amplification of turbulent vorticity caused by particle rotation under varying magnetic field frequencies. Although modulating turbulence using light particles is of particular interest to us since such particles interact strongly with turbulent vorticity, tuning particle density to direct them into different regions of the flow can also provide valuable insights and broaden our understanding of turbulence modulation. In addition, our rotation tracking method offers a powerful diagnostic tool for probing rotational intermittency in turbulence, by tracking particles of various densities across regions with different vorticity levels. Furthermore, using particles of different sizes enables us to investigate how turbulence behaves across scales, as particles inherently act as spatial filters of turbulent structures \cite{Naso2010}. To further elucidate the mechanisms underlying this modulation, we plan to integrate Lagrangian Particle Tracking (LPT) techniques \cite{Tan2020} into our experimental framework. This will allow for the simultaneous reconstruction of both translational and rotational motions of particles, offering a more complete picture of particle–turbulence interactions.

\begin{acknowledgments}
The authors thank F. van Uittert, G. Oerlemans, and J. van der Veen for their technical support. We are also grateful to G. Voth for his inspiring contributions to the development of the image processing algorithm. Special thanks go to R. Lavrijsen for his support in the experimental characterization of the magnetic field and the measurement of the magnetic susceptibility of the particles, which were carried out in the NanoAccess Laboratories at TU Eindhoven. This publication is part of the project “Shaping turbulence with smart particles” with Project No. OCENW.GROOT.2019.031 of the research program Open Competitie ENW XL which is (partly) financed by the Dutch Research Council (NWO).
\end{acknowledgments}

\appendix \label{Appendix}
\section{Parameter design of the setup}
\label{Appendix1}

Table \ref{tab:parameter_formula} summarizes the equations used to estimate the properties of the turbulent flow and the magnetic field. Given the radius, separation distance, and rotation frequency of the impellers ($R_\text{d}$, $H_\text{d}$, and $\Omega$), as well as the particle diameter and its anisotropic magnetic susceptibility ($D_\text{p}$ and $\Delta\chi$), the required coil radius and total current ($R_\text{c}$ and $nI$) can be calculated to ensure that the magnetic torque balances the turbulent torque. To support parameter exploration, an Excel spreadsheet is included in the Supplemental Material \cite{SupplementalMaterial2025}, enabling users to input various parameter values and observe the resulting adjustments needed to maintain torque balance.

\begin{table}[H]
\caption{\label{tab:parameter_formula}Equations for estimating parameters of the setup with known $R_\text{d}$, $H_\text{d}$, $\Omega$, $D_\text{p}$, and $\Delta\chi$.}
\begin{ruledtabular}
\begin{tabular}{p{0.5\linewidth} p{0.5\linewidth}}
\centering \textbf{Parameter} & \centering \textbf{Equation} \tabularnewline
\hline
\centering $\varepsilon$ (W/kg) & \centering $  \left(C_\varepsilon R_\text{d}^3\Omega^3\right)/H_\text{d}$ \tabularnewline
\centering $\eta$ (m) & \centering $\left( \nu^3 H_d / \left({C_\varepsilon R_d^3\Omega^3}\right)\right)^{1/4}$ \tabularnewline
\centering $\tau_\eta$ (s) & \centering $\left( \nu H_d / \left(C_\varepsilon R_d^3\Omega^3 \right) \right)^{1/2}$ \tabularnewline
\centering $\text{Re}_\lambda$ & \centering $\sqrt{15/\nu} R_\text{d}^{2/3} \varepsilon^{1/6}$ \tabularnewline
\centering $\delta \omega$ (1/s) & \centering $\sqrt{14/45} \varepsilon^{1/3} r^{-2/3},\ r = D_\text{p}$ \tabularnewline
\centering $T_{\text{st}}$ ($\text{N}\; \text{m}$) & \centering $\pi \rho \nu D_\text{p}^3 \delta\omega$ \tabularnewline
\centering $T_\text{m}$  ($\text{N}\;\text{m}$) & \centering $\pi \Delta \chi D_\text{p}^3 |\bm{B}^2|/(6\mu_0)$  \tabularnewline
\centering $|\bm{B}|$ (T) & \centering $\left(4/5\right)^{(3/2)}\left(\mu_0nI\right)/R_\text{c}$ \tabularnewline
\end{tabular}
\end{ruledtabular}
\end{table}

\section{Characterization of the Magnetic Properties of the Particles}
\label{Appendix2}

\begin{figure}[htbp]
    \centering
    \includegraphics[width=1\textwidth, angle=0]{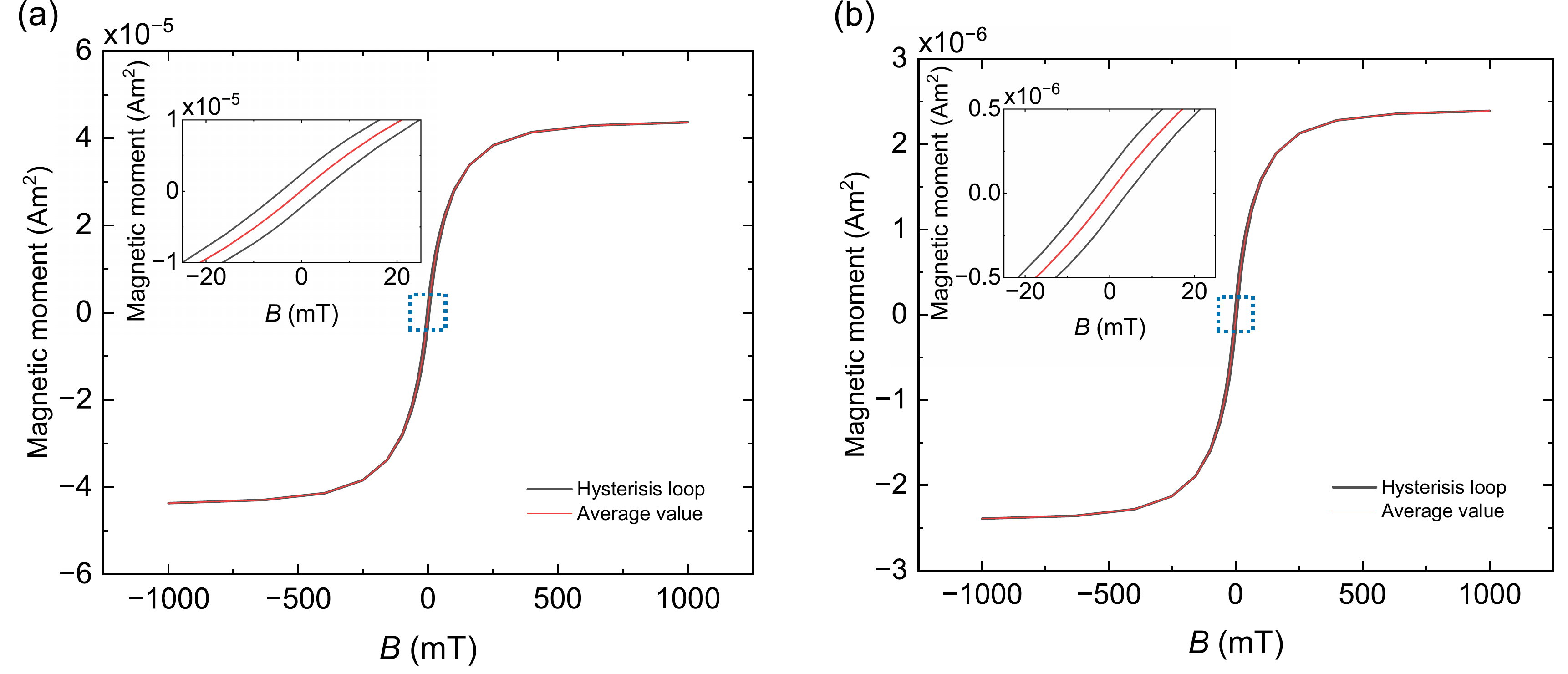 } 
    \caption{(a) The hysteresis loop (black line) shows the magnetic response of a group of 20 randomly oriented magnetic particles subjected to a one-dimensional magnetic field with a flux density ranging from –1 T to 1 T. The red line represents the average of the ascending and descending branches of the loop. The slope of this averaged curve is proportional to the magnetic susceptibility $\chi$. (b) The measurement of $\chi_\parallel$, obtained by aligning the induced magnetic moment of the particle with the applied magnetic field.}
    \label{fig:Chi}
\end{figure}

The average volume magnetic susceptibility $\chi_V$ of the magnetic particles is measured using group samples, each consisting of 20 randomly oriented particles. The measurement is performed under a one-dimensional magnetic field with a flux density ranging from –1 T to 1 T, meaning the direction of the magnetic field is reversed with a maximum magnitude of 1 T in either direction. During the measurement, the particles remain stationary and do not rotate. Figure \ref{fig:Chi}a shows the measured hysteresis loop for one sample group. The particles reach magnetic saturation rapidly when the magnetic field exceeds approximately 500 mT. To eliminate any residual magnetization, the measurement begins at –1 T, increases to 1 T, and then returns to –1 T. The true magnetic moment induced by the external field is obtained by averaging the magnetization values from the ascending and descending branches of the hysteresis loop. The slope of this averaged curve is proportional to the volume magnetic susceptibility $\chi_V$, which is measured to be $0.136 \pm 0.046$.

To measure the anisotropic magnetic susceptibility $\Delta\chi = \chi_\parallel - \chi_\perp$, a magnetic field of 1 T is first applied to allow the particle to freely rotate and align its magnetic moment $\bm{m_\text{p}}$ with the magnetic field $\bm{B}$. Once aligned, the hysteresis loop is measured along this direction to obtain the parallel susceptibility $\chi_\parallel$. Subsequently, $\chi_\perp$ is measured by applying the magnetic field in the direction perpendicular to the aligned magnetic moment. An example of the measurement of $\chi_\parallel$ is shown in Fig. \ref{fig:Chi}b, yielding $\chi_\parallel \approx 0.15$. The corresponding perpendicular susceptibility of this particle is $\chi_\perp \approx 0.14$, resulting in an anisotropic magnetic susceptibility of $\Delta\chi \approx 0.01$.

\bibliography{Ref}

\end{document}